# Silicon Nanostructures for Biosensing: From Field-Effect Transistors to Photonic Resonators, and the Long Road to the Clinic

Ang Liu[1,*], Jun Cao[2], Zhihao Sun[3], Jingsong Shang[4]

[1] Independent Researcher, Los Angeles, CA, USA

[2] Independent Researcher, Fremont, CA, USA

[3] Independent Researcher, Boston, MA, USA

[4] Independent Researcher, Boston, MA, USA

[*] Author to whom correspondence should be addressed: Ang Liu, Email: aliu12@bu.edu

## Abstract

Silicon has a unique combination of properties that makes it one of the best material choices for biosensor platforms: it is inexpensive, its native oxide is atomically smooth, its fabrication processes are CMOS-compatible and have been refined for more than three decades, and it can support many transduction mechanisms in biosensor design. Over the past thirty years, researchers and engineers have used silicon nanostructures to produce ion-sensitive transistors, ultrasensitive nanowire field-effect biosensors, refractive-index-based porous silicon films, microring photonic resonators, suspended cantilevers, luminescent quantum dots, and solid-state nanopores. These device families have demonstrated successful sensing capabilities at the single-molecule, single-virus, or sub-femtomolar level under laboratory conditions; however, they have rarely been widely deployed in clinical assays. This gap is mainly caused by several well-characterized bottlenecks: for nanowire BioFETs, device variability and Debye screening; for porous silicon, fouling, pore wetting, and surface stability; for silicon photonics, thermal drift, spectral readout, and packaging; and across all platforms, calibration, reproducibility, and validation in real biofluids. In this review, we trace the development of silicon biosensors from their early stages to their current state, search and organize the literature focusing on the three most mature platforms and a set of emerging directions, summarize and compare the performance and bottlenecks of different platforms, and argue that progress over the next decade will come primarily from integrated readout, interface engineering, and systematic benchmarking rather than from the discovery of new silicon nanostructures.

**Keywords:**

*silicon nanostructures; biosensors; silicon nanowire FET; porous silicon; silicon photonics; microring resonator; solid-state nanopore; SERS; label-free detection; clinical diagnostics*

# 1. Introduction

Over the past six decades, silicon – the backbone of microelectronics – has emerged as a workhorse of biosensing. In 1970, Bergveld introduced what became known as the ion-sensitive field-effect transistor (ISFET), demonstrating that extracellular ion fluxes from a frog nerve could be recorded using a gateless MOS transistor [1]. The ISFET, however, was not conceived as a biosensor in the modern sense; rather, it represented an early experiment in placing the gate interface of a transistor directly in contact with an ionic electrolyte and observing the resulting response.

In retrospect, what emerged from this experiment was the foundation of the broader BioFET field. Bergveld and colleagues' later retrospective [2], illustrated in Figure 1(A), together with the subsequent BioFET literature [3, 4], makes clear that silicon's dominance in solid-state biosensing has never depended on a single electrical or optical property in isolation. Instead, its strength lies in the ability of silicon-based biosensors to transduce biological recognition events into signals that can be read, processed, and scaled using semiconductor technology.

The modern face of this story is silicon nanostructures. As the same processes that make CMOS logic possible were adapted to define nanowires, nanoribbons, nanopores, and slot-waveguides on semiconductor wafers, the central question shifted from "Can silicon transistors sense ions?" to "How can a biochemical event be translated most effectively into a measurable electrical, mechanical, or optical signal?"

As emphasized in critical reviews from 2021 [5] and 2025 [6], silicon-based biosensors are now best understood as a family of devices spanning charge-based, mechanical, refractive-index, fluorescence, and single-molecule readouts on a common materials platform. This family includes silicon nanowire BioFETs [7], porous silicon optical and electrochemical structures, silicon and silicon-nitride photonic resonators, microcantilevers, quantum dots, and solid-state nanopores. What unites these devices is not a single transduction mechanism, but a common manufacturing language. Morales and colleagues provided a useful practical guide for selecting appropriate biorecognition partners for specific analytes and biosensor formats [8]. The influential 2008 review by Fan and co-workers [9] further lays out the broader case for nano-enabled, label-free optical biosensing, both silicon-based and otherwise.

## 1.1. From discovery to platform

Looking back at the history of silicon biosensors, numerous individual experiments have propelled the field forward by revealing new possibilities and opening previously unexplored directions. In 1990, Canham observed that anodization could render silicon highly porous, giving rise to strong visible photoluminescence [10], as illustrated in Figure 1(C). This serendipitous finding transformed what might have been viewed as a fabrication anomaly into a new class of sensing material. To make porous silicon suitable for biosensing, the material was subsequently functionalized through stable covalent attachment chemistries, including Stewart and Buriak's photopatterned hydrosilylation [11] and Buriak's Lewis-acid-mediated hydrosilylation [12].

On the device side, Cui and colleagues reported in Science in 2001 that bottom-up silicon nanowires could be integrated into transistor architectures and used for highly selective biomolecular detection [17], as illustrated in Figure 1(B). This approach was later translated into a more broadly reproducible methodology through the protocol paper from Patolsky's group [13]. In parallel, Park and co-workers demonstrated top-down nanowire fabrication [14], showing that the same device geometry could be achieved at the wafer level through scalable processing. Subsequent studies by Tintelott et al. [15] and Tran et al. [16] further catalogued the variability that the field continued to inherit from these fabrication choices, particularly in CMOS-compatible process flows.

A porous silicon optical interferometric biosensor was demonstrated by Lin and colleagues in 1997 [36], as illustrated in Figure 3(C). This work, together with the Protein A binding experiment reported by Dancil et al. two years later [37] and the encoded photonic crystal screening approach developed by Cunin et al. [38], established porous silicon as a refractive-index transducer. Pacholski's RIFTS strategy subsequently showed how to extract a clean optical signal from a multilayer stack [39], and the field was further consolidated in the 2009 review by Jane et al. [40], as illustrated in Figure 3(A).

In parallel, integrated silicon photonics produced its own series of breakthroughs. Prieto et al.'s interferometric nanodevice on silicon technology [83] provided an early demonstration, while the SOI microring resonator reported by De Vos et al. [69], as illustrated in Figure 1(D), the slot-waveguide concept demonstrated by Barrios et al. [70, 71], and the slot-waveguide ring developed by Claes and co-workers [72] collectively opened the modern era of silicon photonic biosensing. The photonic-wire array reported by Densmore et al. [73], the quantitative serum biomarker measurements demonstrated by Washburn et al. [74], and the scaled microring array with high-speed scanning instrumentation developed by Iqbal et al. [75] advanced these structures from physics demonstrations toward practical bioassays. These developments were later connected to broader lab-on-a-chip integration in the 2012 review by Estevez et al. [76].

On the mechanical side, as illustrated in Figure 1(F), a defining demonstration was the prostate-specific antigen measurement using microcantilevers reported by Wu and colleagues in 2001 [86]. Subsequent contributions by Raiteri et al. [85], Lavrik et al. [87], Hansen and Thundat [88], Fritz [89], and Arlett et al. [90] helped establish the canonical framework of microcantilever sensing. More recently, monolithic SOI CMOS cantilevers have been brought back into focus by Liu et al. [91].

Solid-state nanopores followed a parallel trajectory, beginning with the ion-beam-sculpted apertures demonstrated by Li et al. [57], the nanometer-precision nanopores reported by Storm et al. [58], as illustrated in Figure 1(E), the single-stranded DNA detection demonstrated by Fologea et al. [59], and the first-pass sequencing-style signals reported by Yan and Xu [60]. Foundational perspectives by Dekker [61], Branton [62], Wanunu [63], and Miles [64] further defined the conceptual and methodological basis of the field. The analytical vocabulary of nanopore sensing was later formalized by Plesa and Dekker [65], while machine-learning-based protein identification using nanopores [66] and localized nanopore fabrication with femtosecond laser pulses [67] expanded the field's analytical and fabrication capabilities. Taken together, Figure 1 summarizes the major device families and their canonical demonstrations.

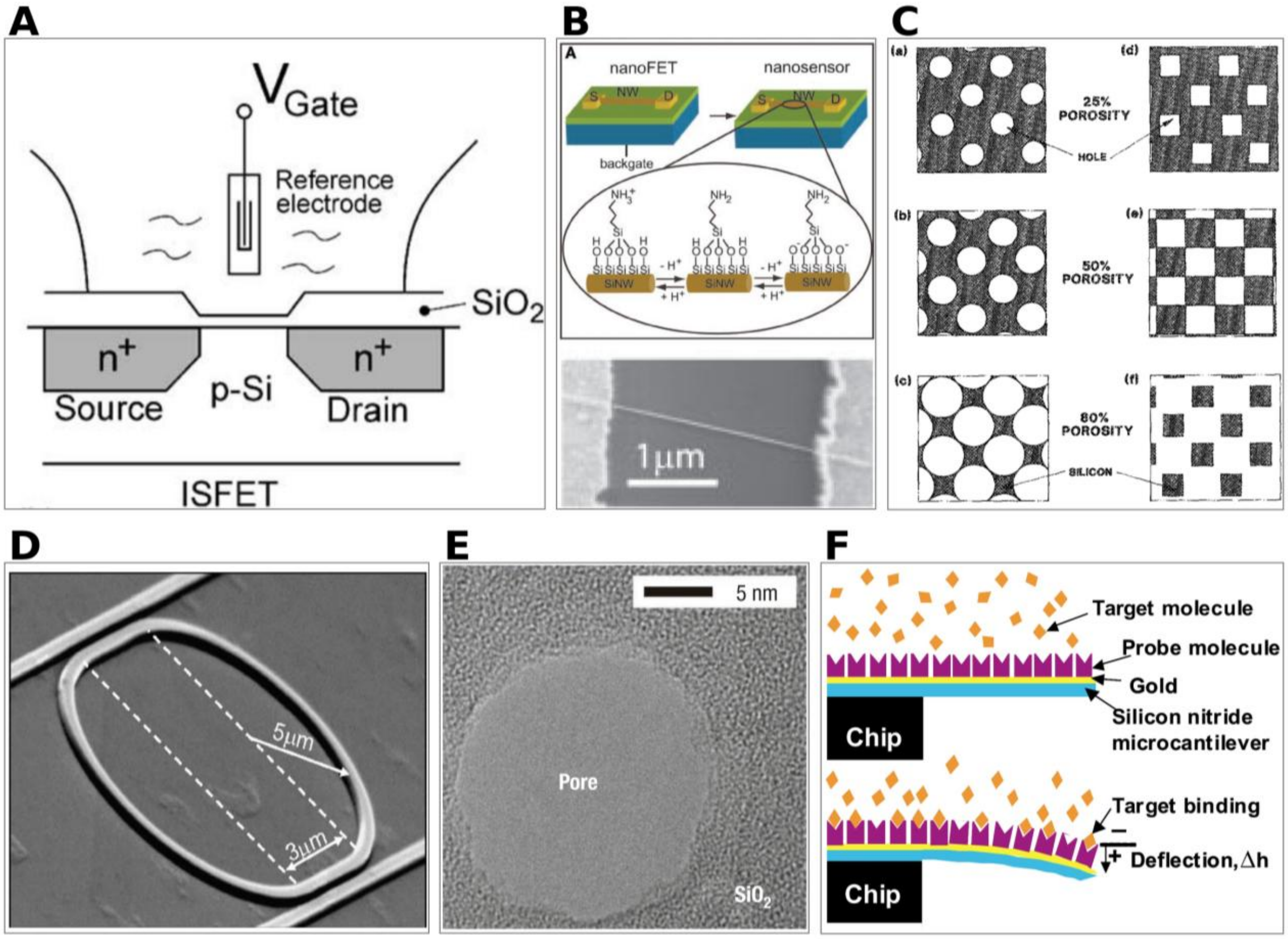

**Figure 1.** Historical and mechanistic landscape of silicon nanostructure biosensors. (A) Schematic of the ion-sensitive field-effect transistor (ISFET), the conceptual ancestor of BioFETs; adapted from Bergveld [2] with permission from Elsevier, copyright 2003. (B) Device schematic and scanning electron micrograph of a bottom-up silicon nanowire field-effect biosensor; adapted from Cui et al. [17] with permission from AAAS, copyright 2001. (C) Idealized plan view of an electrochemically anodized porous silicon film; adapted from Canham [10] with permission from AIP Publishing, copyright 1990. (D) SEM picture of a silicon-on-insulator microring resonator cavity used for label-free refractive-index sensing; adapted from De Vos et al. [69] with permission from Optica Publishing Group, copyright 2007. (E) TEM image of a solid-state nanopore drilled in a silicon-oxide membrane; adapted from Storm et al. [58] with permission from Springer Nature, copyright 2003. (F) Schematic of microcantilever arrays used for label-free PSA detection; adapted from Wu et al. [86] with permission from Springer Nature, copyright 2001.

## 1.2. Transduction mechanisms, briefly

For the scope of this review, we focus on four physical transduction mechanisms. In silicon BioFETs, including nanowires, nanoribbons, and their nanoplate descendants, the channel potential of a subthreshold transistor can be modulated by the charge of an adsorbed biomolecule, producing an exponential change in the source-drain current. The fundamental limits of gating sensitivity were clarified by the dual-gated analysis of Knopfmacher and colleagues [22], and these limits have been addressed through biorecognition-layer engineering by Elnathan et al. [24] and high-frequency operation beyond the Debye length by Kulkarni, Zhong, and co-workers [25]. The resulting current shifts were linked to biophysical affinities and binding kinetics by Duan et al. [26], while detection in high-ionic-strength solutions was demonstrated by Gao and co-workers [27].

In porous-silicon photonic films, following the RIFTS framework established by Pacholski and co-workers [39], binding events alter the effective optical thickness of a Fabry–Pérot or rugate-encoded multilayer, thereby shifting interference fringes or reflectance peaks. In silicon and silicon-nitride photonic resonators, binding events change the local refractive index and shift the resonance wavelength of a microring or photonic-crystal cavity; these shifts can be read out using on-chip filter banks [81], tunable lasers, or integrated spectrometers [79, 82]. In solid-state nanopores, analyte translocation modulates the baseline ionic current, producing blockade-amplitude and dwell-time signatures that can be classified electrically [59, 63] and, increasingly, using machine-learning approaches [66].

$$\lambda_D = \left(\frac{\varepsilon_r \varepsilon_0 k_B T}{2 N_A e^2 I}\right)^{1/2} \tag{1}$$

where $\lambda_D$ is the Debye screening length, $\varepsilon_r$ is the relative permittivity of the electrolyte, $\varepsilon_0$ is the vacuum permittivity, $k_B$ is the Boltzmann constant, T is the absolute temperature, $N_A$ is Avogadro’s number, e is the elementary charge, and I is the ionic strength of the solution. In physiological buffers ($I \approx 150$ mM) at 298 K, $\lambda_D \approx 0.8$ nm. This value represents a practical screening limit and has motivated much of the BioFET literature on linker chemistry and high-frequency operation [22, 24, 25].

$$I_D \propto \exp\left(\frac{e\Delta\Psi_0}{n k_B T}\right) \tag{2}$$

where $I_D$ is the source-drain current of a SiNW or nanoribbon FET operating in the subthreshold regime, $\Delta\Psi_0$ is the surface-potential change induced by the charge of the adsorbed analyte, and n is the subthreshold slope factor, typically in the range of 1–2. BioFETs derive their sensitivity from this exponential dependence; however, drift at the gate-electrolyte interface is also amplified. For this reason, interface engineering and rigorous calibration are critical to reliable BioFET operation [24, 28, 29].

$$\Delta\lambda_{res} = \left(\frac{\lambda_{res}}{n_g}\right) \times \Delta n_{eff} \quad (3)$$

where $\lambda_{res}$ is the resonance wavelength of the cavity, $n_g$ is the group index of the waveguide mode, and $\Delta n_{eff}$ is the change in effective mode index produced by surface-binding events or bulk refractive-index variations. The factor $\lambda_{res}/n_g$ determines the conversion from an index change to a spectral shift. For SOI microrings operating near 1550 nm, this corresponds to a typical bulk sensitivity of approximately 50–200 nm per refractive-index unit (RIU) [69, 73, 74].

$$\frac{\Delta I}{I_0} \approx -\frac{d_a^2 L_a}{d_p^2 L_p} \quad (4)$$

where $\Delta I$ is the change in pore current during analyte translocation, $I_0$ is the open-pore baseline current, $d_a$ and $L_a$ are the equivalent diameter and effective length of the analyte segment within the pore, respectively, and $d_p$ and $L_p$ are the diameter and effective length of the silicon nitride pore, respectively. Introduced by Plesa and Dekker [65], this formula provides the geometric basis for dwell-time and blockade-amplitude analysis and has been extended in the machine-learning-based protein-identification work of Dutt and colleagues [66].

The four mechanisms – charge-based transduction, refractive-index transduction, single-molecule pore-current transduction, and luminescent transduction – each aligns with silicon's established manufacturing strengths. Charge-based transduction benefits from well-isolated, long, atomically clean channels, which are naturally enabled by silicon-on-insulator platforms. Refractive-index transduction requires high quality factors, small mode volumes, and strong vertical dielectric contrast, all of which are core advantages of silicon-on-insulator photonics and have been further developed through telecom-driven integration. Silicon nitride, in turn, provides thin, robust, electrically insulating membranes that are well suited for single-molecule pore-current transduction. Finally, for luminescent transduction, silicon can serve as an effective host for quantum-confined emitters [54, 55], for example, biodegradable luminescent porous-silicon nanoparticles have been used for in vivo imaging with remarkably high time-gated contrast [56].

## 1.3. Performance metrics and why we should not compare on LoD alone

It is tempting to rank silicon biosensors solely by their lowest reported limits of detection, but such comparisons can be misleading. A nanowire array may detect femtomolar concentrations of microRNA in a buffer, yet perform poorly in undiluted plasma. Similarly, a photonic microring may resolve single-nucleotide variants under controlled conditions but lose performance as fouling layers accumulate on the waveguide surface. A more rigorous ranking or comparison strategy should therefore account for the limit of detection in a clearly specified matrix, dynamic range, assay time, selectivity relative to competing platforms, multiplexing capacity, drift, sensor-to-sensor variability [28, 29], packaging and instrumentation requirements, and the stability of the underlying chemistry during reuse and storage. This broader perspective has been explicitly emphasized in recent reviews [5, 6, 35, 68], and is further supported by the calibration-first framework introduced by Chen et al. [29], which cautions against limit-of-detection-only comparisons.

Conventionally, the limit of detection is defined by the noise floor and the slope of the calibration curve, as shown in Eq. (5).

$$LoD = \frac{3\sigma_y}{S} \tag{5}$$

where $\sigma_y$ is the standard deviation of the readout noise, expressed in the units of the measured response, such as resonance wavelength, impedance, or drain current, and S is the analytical sensitivity, defined as the slope dy/dc of the calibration curve near the lower end of the dynamic range. The 3σ criterion follows the convention commonly adopted by silicon-biosensor studies [29, 68, 80, 81]. The practical importance of Eq. (5) is that $\sigma_y$ is determined not only by the chip itself, but also by the readout instrument. This dependence is one reason why integrated-readout literature receives substantial emphasis in this review.

# 2. Fabrication of Silicon Biosensing Nanostructures

## 2.1. Silicon nanowires and nanoribbons

There are two dominant fabrication routes, namely bottom-up and top-down approaches, and the difference between them matters more than it may seem. In the bottom-up route, nanowires are grown from gold or other catalyst seeds, harvested from the source wafer, and finally aligned onto a device substrate before electrical contacts are defined. As illustrated in Figure 2(B), Cui and co-workers used this route in their seminal 2001 paper [17] to fabricate silicon nanowires, and their method remains one of the cleanest ways to obtain sub-20-nm-diameter wires with smooth sidewalls. Patolsky's group further developed this approach into a method of fluidic alignment [13], as shown in Figure 2(A), and single-virus electrical detection was enabled by the same crystal-clean morphology [18]. The same chemistry was used by Zheng et al. [19] for their multiplexed cancer-marker arrays. The major challenges of the bottom-up approach for nanowire fabrication are yield and alignment: the wires themselves are excellent, but transforming a full culture dish of wires into a wafer of delicately aligned devices is difficult.

In contrast, the top-down approach starts with a silicon-on-insulator (SOI) wafer, and electron-beam lithography or deep-UV lithography is used to define the patterns of wires or ribbons in the device layer. Stern and colleagues took this approach and made CMOS-compatible nanowire devices for label-free immunodetection [20], and as illustrated in Figure 2(F), Elfström et al. showed that nanoribbons, which are wider than nanowires, are easier to align, contact, reproduce, and apply to many analyte species [21]. Park and co-workers explicitly compared the two methods and showed that top-down fabricated devices can achieve real-time gas and liquid sensing [14], as shown in Figure 2(E). Gao et al.'s work [23] and Zafar's work [28] then showed that wafer-scale CMOS-compatible processing was no longer just an aspiration, although Tintelott and colleagues' study of the process flow for top-down fabrication of SiNWs using Sidewall Transfer Lithography (STL) [15], as illustrated in Figure 2(D), demonstrates process variability and reminds us that, in real-world fabs, device-to-device sensitivity scatter can exceed analyte-driven signals. Tran et al.'s 2018 review [16], as shown in Figure 2(C), summarizes the critical decision-making parameters, including channel doping, oxide thickness, ohmic contacts, and encapsulation chemistry, that determine whether CMOS-compatible nanowire biosensors fabricated in the cleanroom are in good working condition. Representative process flows and fabricated device images from both routes are collectively shown in Figure 2.

There are several additional practical lessons in the fabrication of nanowires and nanoribbons. First, since the depletion region scales similarly across diameters below 50 nm, channel doping matters more than diameter for sensitivity in this size range. Second, the dominant source of 1/f noise is surface roughness at the etched sidewalls, which limits the low-frequency signal-to-noise ratio more than supposedly fundamental thermal limits. Third, operating the device in air or liquid is not simply a packaging detail; it effectively changes the gate environment, the noise floor, and the available electrochemical window. Sun et al.'s 2025 work [32] on the dual detection of HBV-DNA and AFP demonstrates how these choices look in a working device.

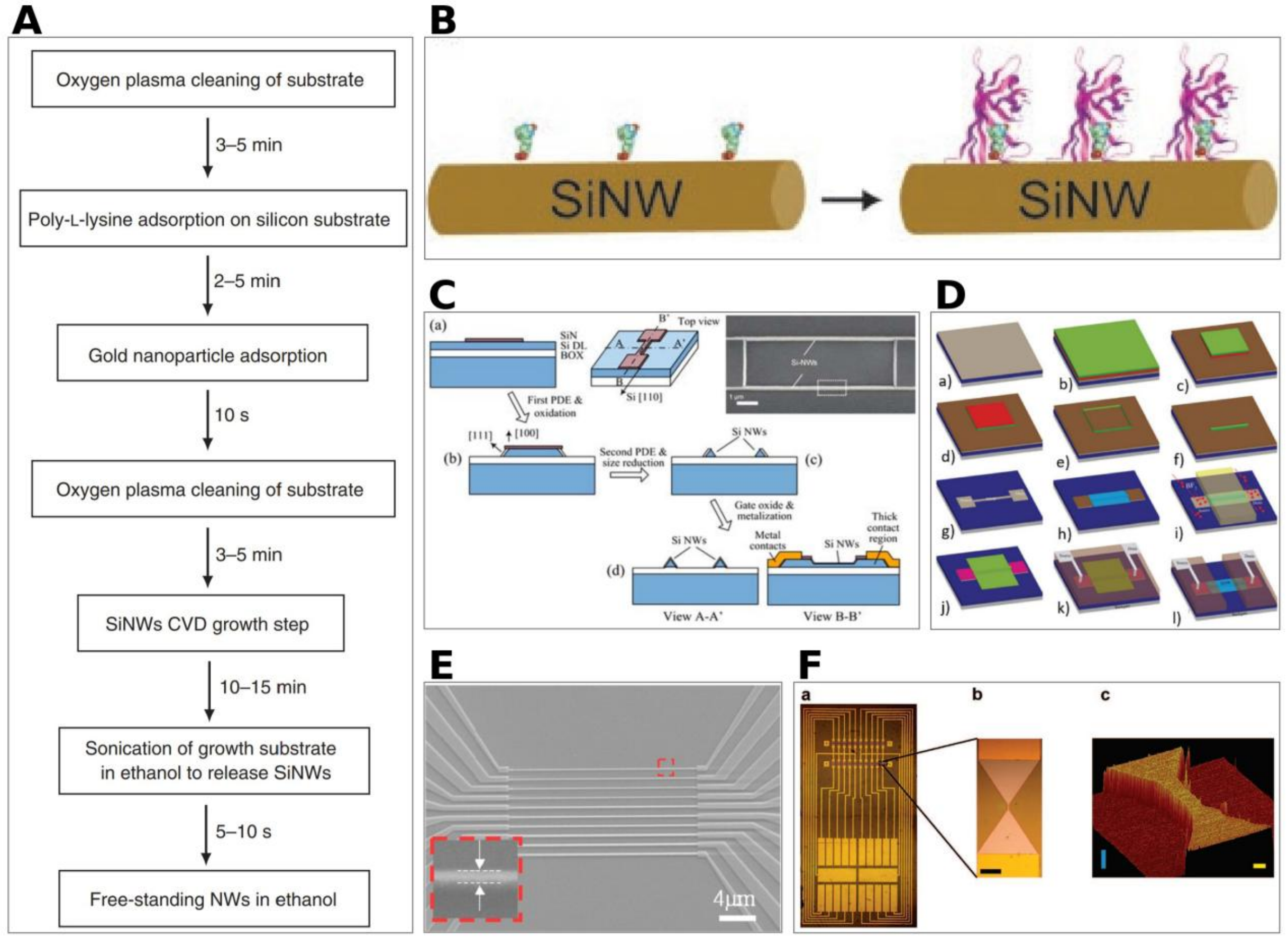

**Figure 2.** Fabrication routes for silicon nanowire and nanoribbon BioFETs. (A) Process flow for bottom-up VLS growth of single-crystal Si nanowires; adapted from Patolsky et al. [13] with permission from Springer Nature, copyright 2006. (B) Schematic illustrating a biotin-modified SiNW (left) and subsequent binding of streptavidin to the SiNW surface (right) on a sensor chip; adapted from Cui et al. [17] with permission from AAAS, copyright 2001. (C) CMOS-compatible wafer-scale nanowire device flow; reproduced from Tran et al. [16] under Creative Commons Attribution License (MDPI, 2018). (D) Process flow of top-down fabrication of SiNWs using Sidewall Transfer Lithography (STL); reproduced from Tintelott et al. [15] under Creative Commons Attribution License (MDPI, 2021). (E) Silicon nanowire sensor arrays fabricated by the top-down nanofabrication process (inset: 50 nm width) defined by electron-beam lithography; adapted from Park et al. [14] with permission from IOP Publishing, copyright 2010. (F) Silicon nanoribbon device with metal contacts and gate; adapted from Elfström et al. [21] with permission from the American Chemical Society, copyright 2008.

## 2.2. Porous silicon

Porous silicon is produced by electrochemical anodization of crystalline silicon in HF-based electrolyte. Since Canham's original publication [10], it has been shown that tuning parameters such as electrolyte composition, current density, doping level, and anodization time can generate pores with diameters ranging from a few nanometers to several hundred nanometers, with porosity values ranging from a few percent to over 80%. By stacking layers with different porosities, one can obtain Fabry-Pérot films, distributed Bragg reflectors, microcavities, and rugate filters with unique optical signatures [36-40]. Cunin utilized photonic-crystal encoding to create barcoded library screens [38], as illustrated in Figure 3(B). The key fabrication parameters include oxidation, porosity gradient, and stabilization. A major limitation is that untreated porous silicon dissolves in aqueous buffers within a few hours; current solutions include silanization, thermal oxidation, and hydrosilylation [11, 12], which can transform this optical material into a usable sensor. Representative geometries and surface-engineering schemes are summarized in Figure 3.

For most porous silicon biosensors, the optical observable is the effective optical thickness (EOT) of the porous layer, as shown in Eq. (6). Binding events can perturb the volume-averaged effective refractive index and thereby shift the fringes in the reflectance spectrum.

$$EOT = 2n_{eff,pSi} \times L_{pSi} \tag{6}$$

where EOT is the effective optical thickness of the porous silicon layer, $n_{eff,pSi}$ is its volume-averaged effective refractive index, which is jointly determined by pore filling, porosity, and adsorbed analytes, and $L_{pSi}$ is the geometrical thickness of the layer. The factor of two accounts for the round trip of normally incident light through the film. As illustrated in Figure 3(D), Pacholski's RIFTS algorithm [39] directly extracts EOT from the Fourier transform of R(1/λ), providing a self-referencing signal that avoids many single-layer artifacts and reduces temperature drift, which affected previous generations of porous silicon biosensors.

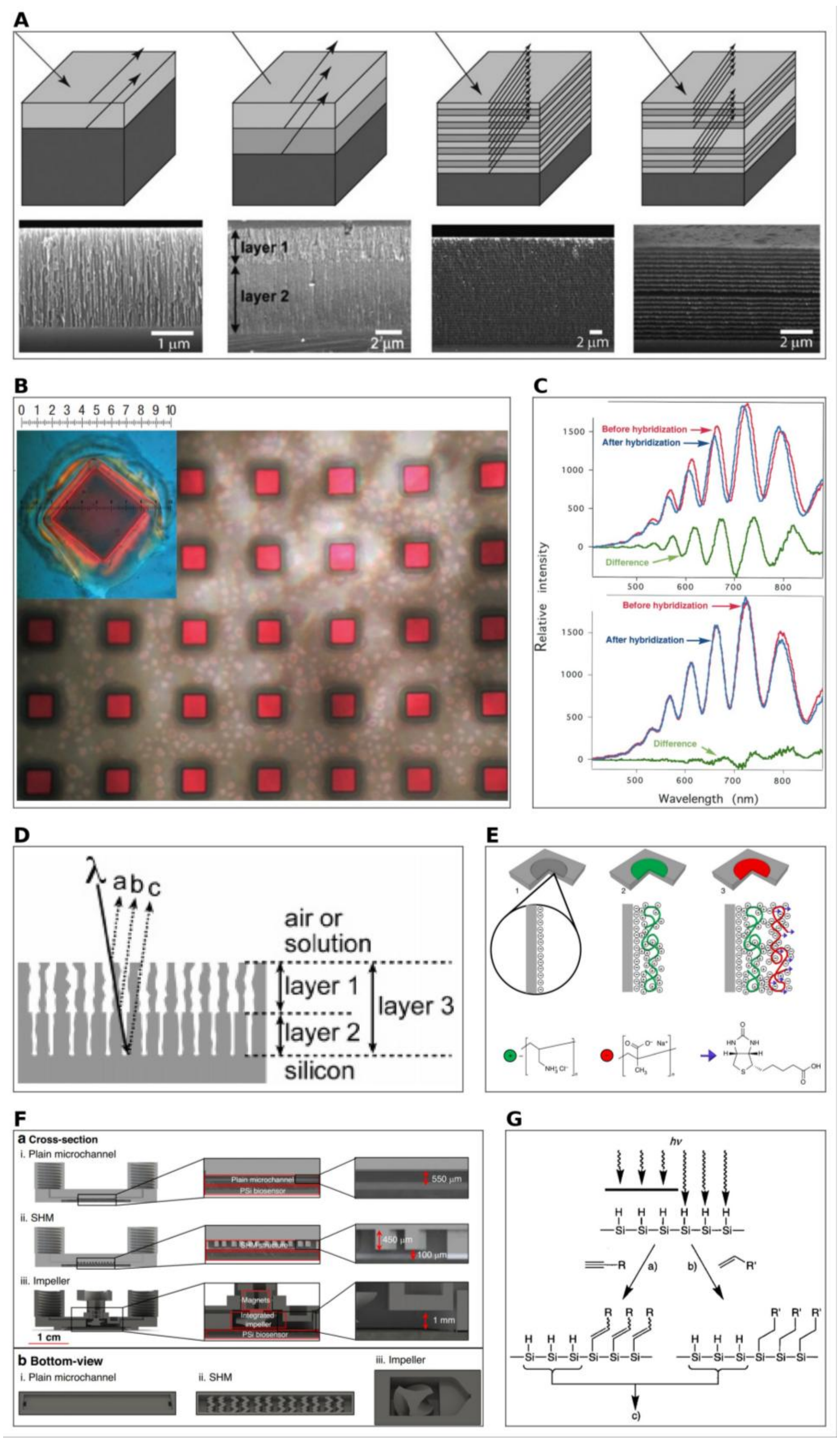

**Figure 3.** Porous silicon fabrication, optical readouts, and surface engineering. (A) Different porous multilayered Si structures; adapted from Jane et al. [40] with permission from Elsevier, copyright 2009. (B) Optical microscope image

of porous-silicon particles photolithographically defined on a silicon wafer; adapted from Cunin et al. [38] with permission from Springer Nature, copyright 2002. (C) Reflectance spectrum of a porous silicon Fabry–Pérot biosensor showing fringe shifts on protein binding; adapted from Lin et al. [36] with permission from AAAS, copyright 1997. (D) Schematic of the reflective interferometric Fourier transform spectroscopy (RIFTS) principle on a porous silicon double-layer device; adapted from Pacholski et al. [39] with permission from the American Chemical Society, copyright 2005. (E) Layer-by-layer biofunctionalization scheme producing stable affinity sensing in complex matrices; reproduced from Mariani et al. [44] under Creative Commons Attribution License (Springer Nature, 2018). (F) Microfluidic integration of a porous silicon membrane optimized against analyte hydrodynamic radius; reproduced from Awawdeh et al. [43] under Creative Commons Attribution License (Springer Nature, 2024). (G) Photopatterned hydrosilylation forming covalent Si–C bonds on porous silicon; adapted from Stewart and Buriak [11] with permission from Wiley-VCH, copyright 1998.

A modern porous-silicon biosensor is never just a bare thin film. Surface stability can be improved using polymeric overlays; the stability of electrochemical platforms in alkaline buffers can be improved through carbon stabilization, as demonstrated by Guo and colleagues [49, 50]; and analyte transport into the deep pore network can be regulated through microfluidic integration. The interplay between microfluidic integration and nanostructure design, as illustrated in Figure 3(F) from Awawdeh et al.'s 2024 paper [43], provides a roadmap for matching pressure drop, pore size, and analyte hydrodynamic radius. Furthermore, the layer-by-layer biofunctionalization scheme proposed by Mariani et al. [44] shows how stable affinity sensing can be engineered in complex matrices, as illustrated in Figure 3(E). When complex matrices need to be probed, flow-through architectures have become a strong alternative to closed-end membranes, as demonstrated in Zhao's study [45].

### 2.3. Silicon photonics

Silicon photonics has the most mature fabrication infrastructure in the biosensor ecosystem, largely because the same SOI process flows are commercially used for data-center transceivers. A typical silicon photonic biosensor wafer contains photonic crystal cavities, Mach-Zehnder interferometers, microring resonators, or Bragg gratings, all defined by 193-nm deep-UV lithography in a 220-nm silicon device layer on a buried oxide. A similar stack, with a SiN core deposited by PECVD or LPCVD, is used in silicon nitride variants. The wavelength-shift readout, typically performed near 1550 nm, translates a binding event into a directly measurable spectral shift, and the quality factors of well-designed rings can exceed $10^5$. To illustrate how the strong field overlap of slot modes can be combined with the resonant amplification of rings, representative architectures are shown in Figure 4, including De Vos's microring [69] in Figure 4(A) and Barrios's slot waveguide [70, 71] in Figure 4(B), which are early canonical devices, as well as Claes and colleagues' slot-waveguide ring [72] in Figure 4(C).

Noise and stability can be controlled by the resonator itself, as well as by fabrication choices such as cladding, coupling, and trench design. As illustrated in Figure 4(D), Densmore et al.'s photonic wire array [73] demonstrates how a single laser can be used to build chip-scale arrays. Subwavelength-grating waveguide sensing, introduced by Wangüemert-Pérez and colleagues [77], showed that conventional strip waveguides can be outperformed by engineered evanescent fields in terms of specific sensitivity, as illustrated in Figure 4(E). Iqbal et al.'s scaled microring arrays [75] paired the chip design with a fast-scanning optical instrument, which is usually the hardest part of the system to scale. More recent fabrication advances have brought the spectrometer onto the chip itself; for example, Yoo et al. [79] integrated a microring with an on-chip near-infrared FT spectrometer, as illustrated in Figure 4(F), and SiN microring resonators have been integrated with an on-chip filter-bank spectrometer [81], as illustrated in Figure 4(G). Both examples demonstrate that the readout, rather than the sensor, was the larger barrier. In contrast, low-cost light-emitting diodes can pump whispering-gallery-mode microdisks and microspheres [84], sacrificing some Q for substantial simplification of the readout.

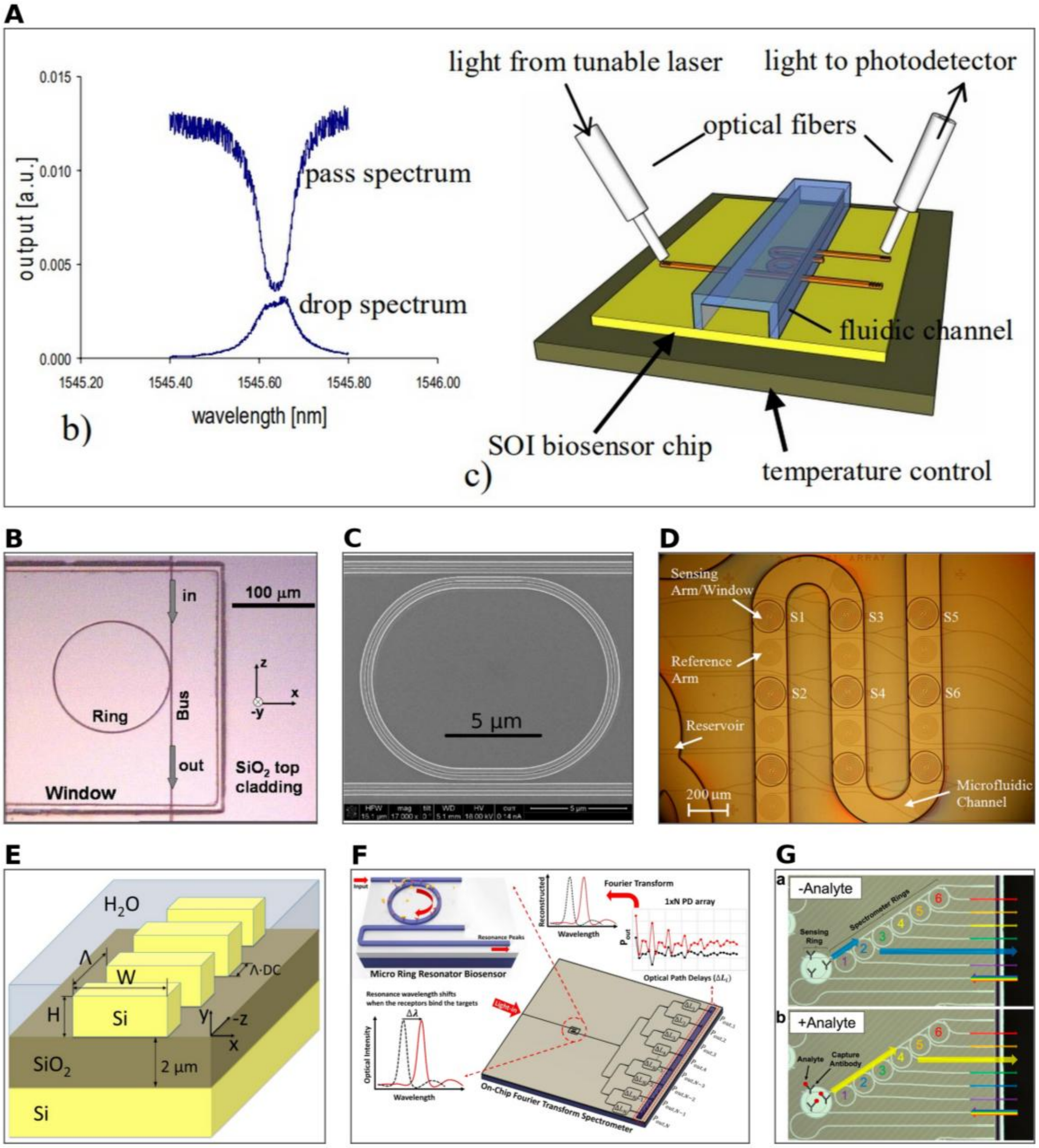

**Figure 4.** Silicon photonic biosensor architectures. (A) Measured transmission spectra and measurement setup for SOI microring resonator; adapted from De Vos et al. [69] with permission from Optica Publishing Group, copyright 2007. (B) Optical microscope top-view photograph of a fabricated slot-waveguide ring resonator. A window is opened in the top over the ring to expose the sensor to different liquids; adapted from Barrios et al. [70] with permission from Optica Publishing Group, copyright 2007. (C) Top-view scanning electron microscope image of the slot-waveguide-based ring resonator; adapted from Claes et al. [72] with permission from IEEE, copyright 2009. (D) Multiplexed silicon photonic wire biosensor array; adapted from Densmore et al. [73] with permission from Optica Publishing Group, copyright 2009. (E) Subwavelength-grating waveguide tailoring the evanescent field for biosensing; adapted from Wangüemert-Pérez et al. [77] with permission from Optica Publishing Group, copyright 2014. (F) Lab-on-a-chip integration of a microring resonator with a near-infrared Fourier-transform spectrometer; reproduced from Yoo

et al. [79] under the arXiv non-exclusive license. (G) Silicon-nitride microring resonator co-integrated with an on-chip filter-bank spectrometer; adapted from Bryan et al. [81] with permission from the American Chemical Society, copyright 2023.

### 2.4. Silicon and silicon-nitride nanopores

The fabrication of solid-state nanopores is conceptually different. Instead of building a resonator or a transistor, the goal is to drill a single nanometer-scale hole through a thin oxide or silicon nitride membrane, after which a transverse electric field is applied to translocate individual biomolecules. Ion-beam sculpting was introduced by Li and colleagues to create nanopores with feedback control and sub-nanometer precision [57], while TEM-based shrinking was used by Storm et al. to refine pore geometry [58]. The most common methods for drilling silicon nitride nanopores are focused-ion-beam drilling and controlled dielectric breakdown. More recently, Leva et al. demonstrated a practical approach for improving yield by fabricating localized nanopores through femtosecond laser exposure followed by dielectric breakdown [67]. The key factors governing system performance include the pore geometry, lipid passivation, surface charge, and noise control [63, 65].

### 2.5. Surface chemistry as a fabrication step

Alongside fabrication, it is worth discussing surface chemistry in depth, as it often determines whether a fabricated sensor functions reliably on the bench. There are three dominant approaches. On oxidized silicon or silicon nitride, organosilanes, such as APTES or GOPS, provide amine or epoxy handles for biomolecule binding. On oxide-free silicon, particularly porous silicon, Stewart and Buriak's photopatterned hydrosilylation [11], as illustrated in Figure 3(G), and Buriak's Lewis-acid-mediated approach [12] create direct Si-C bonds that resist hydrolysis. In addition, for affinity engineering against the Debye limit, Elnathan and colleagues' biorecognition-layer engineering [24] demonstrated that the analyte's charge can be effectively brought within the Debye length through appropriate linker chemistry. For photonic devices, Puumala et al.'s 2023 review paper demonstrates this point in the context of biofunctionalization of multiplexed silicon photonic biosensors [80]; for porous silicon, Mariani and colleagues' Nature Communications report [44] demonstrates the same principle. Zwitterionic peptides, antifouling polymers, and orientation-controlled antibody immobilization are therefore increasingly considered part of the fabrication recipe rather than afterthoughts.

# 3. Results, Applications, and Discussion

## 3.1. Silicon nanowire and nanoribbon BioFETs

Of all silicon-based biosensing technologies, BioFETs have become one of the most widely used platforms for biomarker detection due to their direct conversion of interfacial potential change to electrical signals in a label-free and real-time manner. The silicon nanowire device reported by Cui and co-workers in 2001 [17] established an influential sensing scheme, showing that a nanoscale semiconductor channel could be used as a sensitive transducer for biological binding events. Subsequent studies extended this concept from proof-of-concept sensing to a broader range of biologically relevant targets and device formats. Patolsky and colleagues demonstrated the detection of discrete single-virus binding events [18] as illustrated in Figure 5(A), while Zheng et al. showed that nanowire arrays could be used as a multiplexed cancer-marker detector and reported simultaneous readouts of PSA, PSA-α1-antichymotrypsin, CEA, and mucin-1 at femtomolar concentrations, as illustrated in Figure 5(D). In parallel, Stern et al. successfully developed CMOS-compatible immunodetection [20], as illustrated in Figure 5(B), then tied this performance to a fab-compatible flow, and Gao and co-workers further expanded the platform towards label-free nucleic-acid detection at femtomolar levels [23] and transistor-based biodetection in high-ionic-strength solutions [27]. These representative studies, summarized in Figure 5, demonstrate the versatility of silicon-based BioFETs as a biosensing platform and set the stage for more detailed studies of their sensing mechanisms and performance limitations.

The broad development of silicon BioFETs has therefore been accompanied by extensive efforts to understand the underlying sensing mechanisms and to overcome key limiting factors. Knopfmacher et al. explained qualitatively the device's pH sensitivity with the dual-gated Nernst analysis, as shown in Eq. (7).

$$\frac{\Delta\Psi_0}{\Delta pH} = -\alpha \times \frac{2.303 k_B T}{e} \tag{7}$$

where $\Delta\Psi_0/\Delta pH$ is the change in the gate-oxide surface potential per unit change in bulk pH, and the dimensionless sensitivity parameter α (with $0 \leq \alpha \leq 1$) captures the buffer capacity of the oxide; $\alpha = 1$ corresponds to a perfectly Nernstian response. At 298 K the theoretical Nernst limit ($\alpha = 1$) is −59.2 mV/pH. And they further proposed a potential way of going beyond this Nernst limit with the help of a backgate [22]. Duan et al.'s work [26] presented a clear picture of the protein-binding affinities and kinetics and moved the sensing results of silicon nanowire FETs from a qualitative stage to a more quantitative stage. Zafar and colleagues' demonstration of SiNW-FETs with minimal sensor-to-sensor variation [28], as illustrated in Figure 5(F), showed how careful fab discipline can bring variability under control, and Chen et al.'s calibration strategy [29] formalized this. On the other hand, to overcome the Debye screening effect,

Kulkarni and Zhong [25], as illustrated in Figure 5(C), showed that the screening length effectively grows because counterions cannot follow the analyte's charge fluctuations at high frequencies. Elnathan and colleagues [24], as illustrated in Figure 5(E), demonstrated how a cleaved antibody molecule helps with settling the target analytes sitting inside the screening length to improve silicon BioFETs' sensitivity.

Building on these advances in analyte coverage and mechanistic understanding, recent studies on silicon-based BioFETs have shifted focus toward more complex biological targets, hybrid recognition schemes, and scalable diagnostic sensing platforms. Qin et al. reported successful detection of exosomal-membrane-proteins in a highly sensitive and selective manner using a silicon-based FET biosensor [30]. Li and colleagues' CRISPR/Cas12a-functionalized SiNW-FET for pathogen nucleic-acid detection [31], as illustrated in Figure 5(G), provides a good example of using cleaved receptors to overcome the screening effect. Sun et al. demonstrated an array sensor for dual detection of HBV-DNA and AFP [32]. And at a larger integration scale, Rothberg and coworkers developed the Ion Torrent platform [33], as illustrated in Figure 5(H), and Toumazou and colleagues reported a pH-sensing semiconductor system for simultaneous DNA amplification and detection [34]. These examples show that FET-based sensing principles have a great potential for scaling from a single sensor to highly integrated semiconductor platforms with millions of sensors.

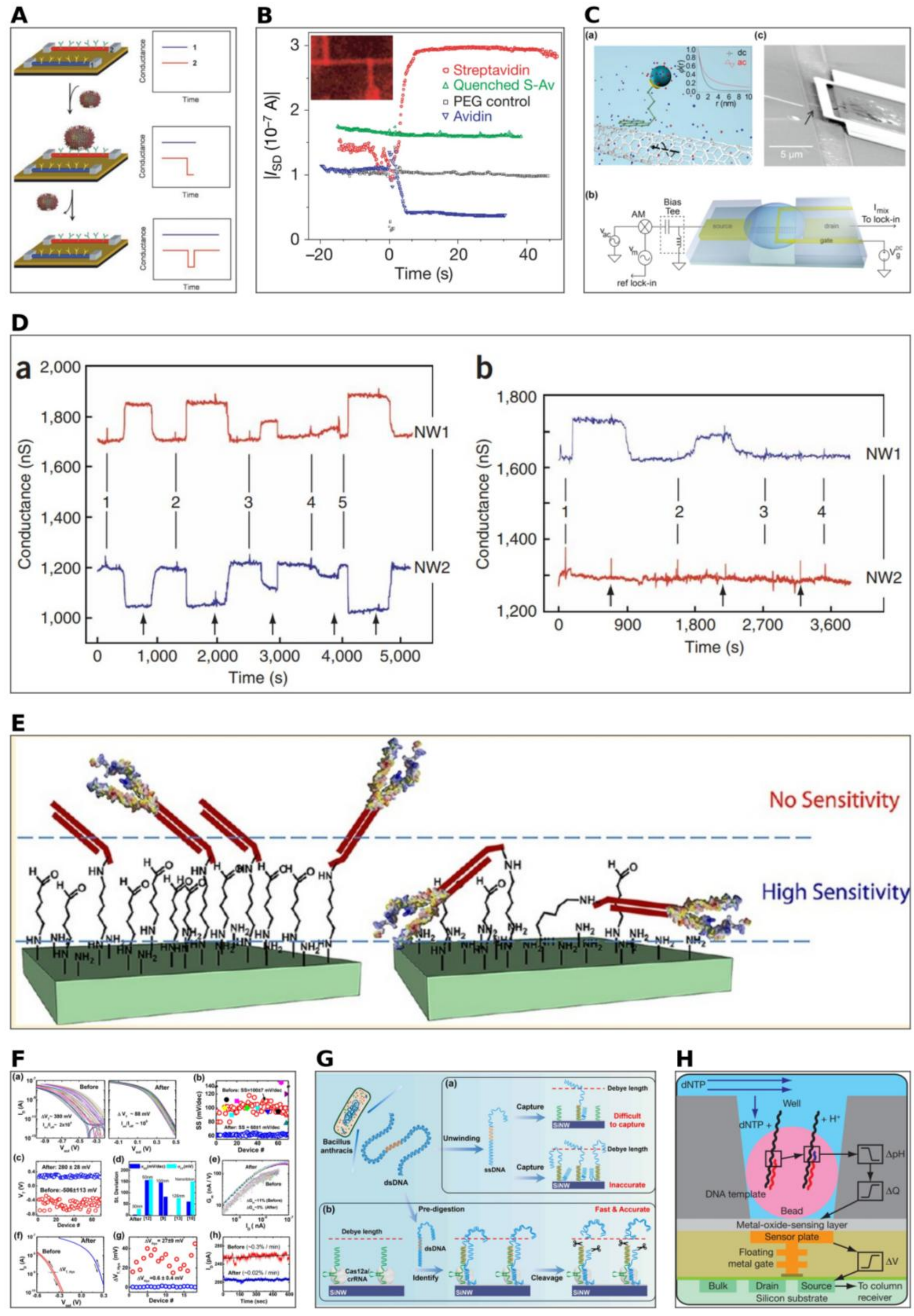

**Figure 5.** Silicon nanowire BioFET detection demonstrations across analyte classes. (A) Conductance traces showing real-time electrical detection of single virus particles; adapted from Patolsky et al. [18] with permission from PNAS,

copyright 2004. (B) Label-free immunodetection of protein markers on a CMOS-compatible nanowire; adapted from Stern et al. [20] with permission from Springer Nature, copyright 2007. (C) High-frequency operation extending sensitivity beyond the Debye screening length; adapted from Kulkarni and Zhong [25] with permission from the American Chemical Society, copyright 2012. (D) Multiplexed simultaneous detection of multiple cancer biomarkers in a single chip; adapted from Zheng et al. [19] with permission from Springer Nature, copyright 2005. (E) Biorecognition layer engineering bringing analyte charges inside the Debye screening length; adapted from Elnathan et al. [24] with permission from the American Chemical Society, copyright 2012. (F) Distribution of device sensitivities across an array, illustrating the variability gains achievable with strict process control; adapted from Zafar et al. [28] with permission from the American Chemical Society, copyright 2018. (G) CRISPR/Cas12a-functionalized SiNW-FET for pathogen nucleic acid detection; adapted from Li et al. [31] with permission from Elsevier, copyright 2025. (H) Ion Torrent semiconductor chip enabling non-optical genome sequencing on a CMOS imager-style array; adapted from Rothberg et al. [33] with permission from Springer Nature, copyright 2011.

### 3.2. Porous silicon biosensors

Unlike SiNW-FET based biosensors where significant effort has been made to extend the categories of the target analyte, the porous silicon biosensors have followed a different development path. Studies on porous silicon biosensors have been focused on bacterial detection, antibody binding, and more recently electrochemical sensing formats. Benefiting from the porous structure, the device comes intrinsically with a large internal surface area and provides abundant binding sites, as well as a great potential for optical interference, photonic-crystal responses, and electrochemical signal amplification. As an early demonstration, Lin et al. developed the interferometric scheme and showed that biomolecular bindings inside porous silicon could be transduced into the wavelength shift in the optical reflectance spectrum in their 1997 work [36], as illustrated in Figure 6(A). Dancil et al. further demonstrated that protein-A affinity binding could be monitored reversibly using porous silicon optical structures [37], as illustrated in Figure 6(B). Cunin et al. generated encoded porous-silicon photonic crystals [38] which extend the traditional single-sensor readout toward multiplexed optical identification. Pacholski et al. introduced the double-layer interferometers and RIFTS algorithm [39], which were then used as a self-referencing readout that survives temperature drift and optical artifacts. These early developments were later summarized by Jane et al. [40], with representative results shown in Figure 6.

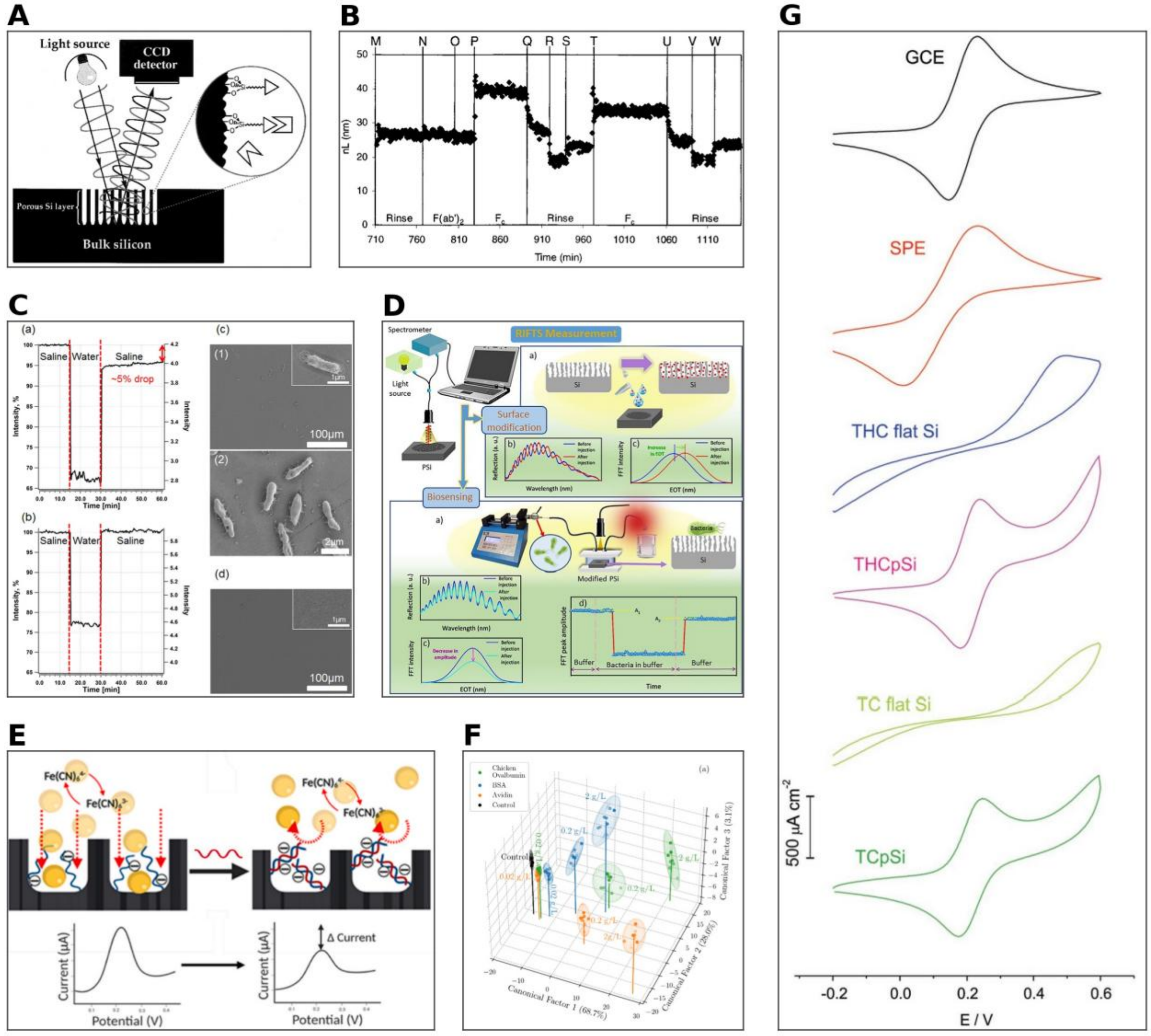

**Figure 6.** Porous silicon biosensing applications. (A) Schematic of porous silicon Fabry–Pérot biosensor for biomolecular binding; adapted from Lin et al. [36] with permission from AAAS, copyright 1997. (B) Reversible binding of IgG to a protein-A-modified porous silicon optical biosensor; adapted from Dancil et al. [37] with permission from the American Chemical Society, copyright 1999. (C) Real-time optical detection of target bacteria in food-industry matrices; reproduced from Massad-Ivanir et al. [41] under Creative Commons Attribution License (Springer Nature, 2016). (D) Label-free real-time bacterial detection using a lectin-coupled porous silicon biosensor; reproduced from Yaghoubi et al. [42] under Creative Commons Attribution License (Springer Nature, 2020). (E) Ultrasensitive carbon-stabilized porous silicon biosensor for bacterial RNA detection; reproduced from Chin et al. [51] under Creative Commons Attribution License (Elsevier, 2024). (F) Capture-agent-free analyte classification using porous silicon arrays and machine learning; reproduced from Ward et al. [46] under the arXiv non-exclusive license. (G) Faradaic electrochemical sensing on a porous silicon nanostructure; adapted from Guo et al. [49] with permission from Wiley-VCH, copyright 2019.

After those early demonstrations, porous silicon biosensors were more extensively explored in application-oriented settings, especially for bacteria detection and sensing in complex environments. Massad-Ivanir et al. [41] demonstrated real-time optical detection of Escherichia coli (E. coli) in complex food industry process water using antibody-functionalized porous silicon optical film, as illustrated in Figure 6(C), and Yaghoubi et al. [42] further developed the E. coli detection with lower-cost lectins and showed the potential of porous silicon for classification-capable microbial sensing, as illustrated in Figure 6(D). These studies highlight the great built-in advantage, the large intrinsic surface area of porous silicon, when used for detecting the whole microbe. In other studies, focusing on the stability and performance of the method, Mariani and colleagues developed a layer-by-layer biofunctionalization protocol for high-stability affinity sensing and showed successful streptavidin detections in saliva [44]. Awawdeh et al. emphasized that the device performance is determined not only by the surface area, but also by pore accessibility, hindered diffusion, probe density, and microfluidic mass transport [43]. Zhao et al. compared the response time of porous silicon membrane sensors and concluded with a six-fold improvement with their flow-through architecture [45]. The Arshavsky-Graham review [35] from 2019 remains an excellent map of where the applications sit relative to the underlying optics.

Recent work has also explored the porous silicon biosensors beyond the conventional optical affinity readout. Ward et al. demonstrated capture-agent-free biosensing using porous silicon arrays with machine learning analysis [46], as illustrated in Figure 6(F). This study suggests that, when many optical responses are collected in parallel, analyte identification can be approached as a pattern-recognition problem. Meanwhile, electrochemical porous silicon has evolved into an important subfield. Setzu et al. [47] and RoyChaudhuri [48] summarized the broader development of this platform, while Guo et al. showed that porous silicon nanostructures can serve as faradaic sensing platforms with amplified electrochemical signals when their conductivity and surface chemistry are properly engineered [49], as illustrated in Figure 6(G). Their later carbon-stabilized layered structure [50] addresses the long-standing stability issue, and Chin et al. demonstrated stable ultrasensitive label-free detection of bacterial RNA gene fragments [51], as illustrated in Figure 6(E). Together, porous silicon biosensors have been proved to be an excellent sensing platform benefiting from their large internal surface area, but this same feature also makes the platform sensitive to fouling, corrosion, pore blockage, and mass-transport limitations.

### 3.3. Silicon photonic biosensors

Silicon photonic biosensors are one of the closest to clinical translation because they leverage existing, highly scalable semiconductor manufacturing processes and provide label-free readout in real-time. A few influential structures have been established early and are still being widely used nowadays. De Vos et al. demonstrated SOI microring resonators for label-free biosensing [69], while Barrios and colleagues [70, 71] developed a slot-waveguide structure that increased the light-matter interaction through stronger field confinement. Claes et al. later combined these concepts in the slot-waveguide ring resonators for detection of avidin [72] and Densmore et al. developed the photonic-wire arrays for simultaneous monitoring of different molecular binding events [73]. These structural-level advances were followed by more clinically oriented demonstrations. Washburn and colleagues [74] showed label-free quantifications of a cancer biomarker in spiked serum using silicon microrings, as illustrated in Figure 7(A), and Iqbal et al. scaled the platform into a 32-channel microring array with high-speed optical scanning instrumentation [75], as illustrated in Figure 7(B). Silicon photonic biosensors have transitioned from individual optical transducers towards broader integrated-optics platforms, as summarized by Estevez et al. [76].

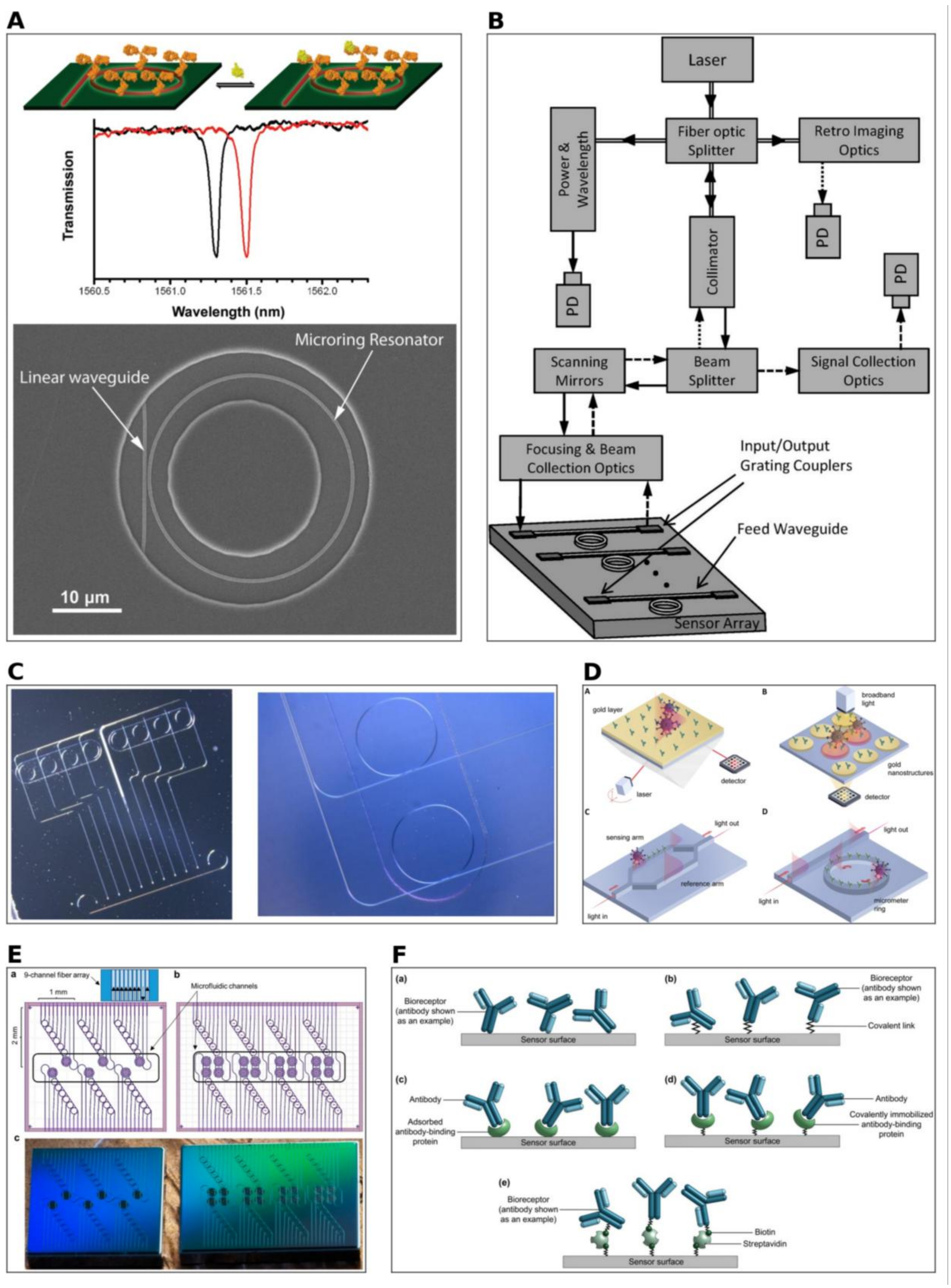

**Figure 7.** Silicon photonic biosensor applications. (A) Schematic diagram illustrating the principle of microring optical resonator biosensing with an SEM image of a microring resonator and linear waveguide; adapted from Washburn et al. [74] with permission from the American Chemical Society, copyright 2009. (B) 32-channel microring array paired with high-speed optical scanning instrumentation; adapted from Iqbal et al. [75] with permission from IEEE, copyright 2010. (C) Fabricated biosensor Photonic Integrated Circuit (PIC) for multiplexed photonic label-free detection of swine viral diseases; reproduced from Gómez-Gómez et al. [82] under Creative Commons Attribution License (MDPI, 2022). (D) Nanophotonic label-free detection strategies for respiratory virus diagnostics, including SARS-CoV-2; adapted from Soler et al. [78] with permission from the American Chemical Society, copyright 2020. (E) SiN microring biosensor integrated with an on-chip filter-bank spectrometer for direct interrogation without a tunable laser; adapted from Bryan et al. [81] with permission from the American Chemical Society, copyright 2023. (F) Comparison of biofunctionalization chemistries on multiplexed silicon photonic biosensors; reproduced from Puumala et al. [80] under Creative Commons Attribution License (MDPI, 2023).

More recent work has shifted the emphasis from the sensing element alone to the complete assay system, including the readout onto the chip and performance in complex matrices. Yoo et al. demonstrated a lab-on-a-chip microring platform integrated with a near-infrared Fourier-transform spectrometer [79], while Bryan et al. integrated SiN microring resonators with an on-chip filter bank spectrometer, as illustrated in Figure 7(E). Both studies highlight a key issue that as silicon photonics sensors mature, the system bottleneck often comes from the instrumentation around the chip. In parallel, Wangüemert-Pérez demonstrated that subwavelength-grating waveguides could be used as a device-level strategy for enhancing evanescent-field interaction and improving refractive-index sensitivity [77]. Kim and Lee showed that active whispering-gallery-mode resonators pumped by an LED could be used as a cost-reducing readout strategy [84]. Application-oriented demonstrations have been expanded quickly. Gómez-Gómez et al. developed SiN ring-resonator biosensors for fast, multiplexed, and label-free detection of swine viral diseases [82], as illustrated in Figure 7(C), while the COVID-19 pandemic further emphasized the need for rapid and multiplexed diagnostics, as discussed in the nanophotonic biosensing review by Soler et al. [78], as illustrated in Figure 7(D). Prieto's earlier integrated interferometric biosensor [83] is one of the field's original blueprints for that kind of clinical device. At the assay level, Puumala and colleagues reviewed biofunctionalization strategies for multiplexed silicon photonic biosensors [80], as illustrated in Figure 7(F), addressing important factors towards a working assay. Recent reviews, including Ramirez-Priego et al.'s work [68], have consolidated the progress of silicon photonic biosensors toward clinical diagnostics.

## 3.4. MEMS cantilever biosensors

Microcantilevers have long occupied a relatively modest position in the broader field of silicon biosensing. Wu et al.'s PSA bioassay [86], illustrated in Figure 8(H), is often regarded as one of the original demonstrations of this approach. The general review by Raiteri et al. [85] further established that label-free receptor-ligand binding could be resolved through the deflection of thin silicon or silicon-nitride cantilevers. The field was subsequently advanced within the surface-stress sensing regime by the chemical and biological sensor review by Lavrik and colleagues [87], the methods paper by Hansen and Thundat [88], and the review by Fritz [89]. Looking ahead, Arlett, Myers, and Roukes [90] argued explicitly that mechanical biosensors may offer comparative advantages when sensitivity per unit mass becomes the relevant figure of merit.

A persistent challenge for mechanical biosensors, however, has been operation and packaging in liquid environments, where viscous damping reduces the quality factor, Q, and suppresses cantilever response. One recent and notable attempt to address system-level integration is the 2023 work by Liu et al. [91], which demonstrated a monolithically integrated microcantilever transducer-amplifier system on partially depleted SOI CMOS, leveraging the maturity of CMOS processing.

As shown by Stoney's relation in Eq. (8), a differential change in surface energy between the functionalized and reference faces of the cantilever is converted by surface-stress microcantilevers into a measurable tip deflection. This expression also explains why microcantilever sensitivity depends strongly on thickness.

$$\Delta z = \frac{3(1-\nu)L^2\Delta\sigma_s}{Et^2} \tag{8}$$

where $\Delta z$ is the deflection of the free end of the cantilever, L is the cantilever length, t is its thickness, E is the Young's modulus, $\nu$ is the Poisson's ratio of the cantilever material, and $\Delta\sigma_s$ is the differential surface stress between the functionalized and reference surfaces. The $L^2/t^2$ scaling, together with the inverse dependence on E, explains why long, thin, low-modulus geometries are generally preferred for microcantilever biosensors. However, operation in aqueous environments also subjects these same geometries to strong viscous damping, which reduces the quality factor, Q. As a result, maintaining adequate Q-factor in liquid has remained one of the central practical limitations in the field [85, 89, 90].

Microcantilever biosensors remain largely a laboratory-stage technology, with limited industrial translation despite nearly three decades of elegant proof-of-concept demonstrations. The recurring challenge is operation in liquid environments. The high mechanical quality factor, Q, that underpins the sensitivity advantage of cantilever-based sensing is precisely the property most strongly degraded by viscous damping in aqueous media.

The most promising path forward may therefore lie not only in improved cantilever materials or dimensional optimization, but also in tighter monolithic integration of the cantilever with CMOS readout circuitry. Such integration could reduce off-chip electronic noise and improve system-level operation in aqueous environments. If mechanical cantilever MEMS biosensors are to compete with FET-based and photonic-resonator biosensors, they will likely do so in niche applications where the cleanest signal arises from analyte mass, rather than charge, as in FETs, or refractive-index contrast, as in photonic resonators.

### 3.5. Emerging directions: nanopores, quantum dots, photonic crystals, black silicon

Several silicon nanostructures are gaining importance as early-stage platforms for biosensor translation. Among them, solid-state nanopores are perhaps the closest to maturity. As illustrated in Figure 8(A), the foundational work by Li [57], Storm [58], and Fologea [59] established the basis of the field. Building on these advances, Yan and Xu [60] and Dekker [61] argued that DNA sequencing could be achieved using this platform, while Branton's perspective on its potential and remaining challenges [62] continues to define the translational checklist for the field. A decade of subsequent refinement has been summarized in Wanunu's review [63] and in the materials-and-methods survey by Miles et al. [64]. The data-analysis tutorial by Plesa and Dekker [65] has become a standard reference for event extraction.

More recently, high-accuracy protein identification enabled by machine-learning-based signal fusion [66], as illustrated in Figure 8(B), together with localized femtosecond-laser-plus-breakdown fabrication of nanopores by Leva and co-workers [67], points toward a future in which silicon-nitride-based nanopores can be fabricated and read out at relatively high throughput. Representative data from these emerging platforms are collected and summarized in Figure 8.

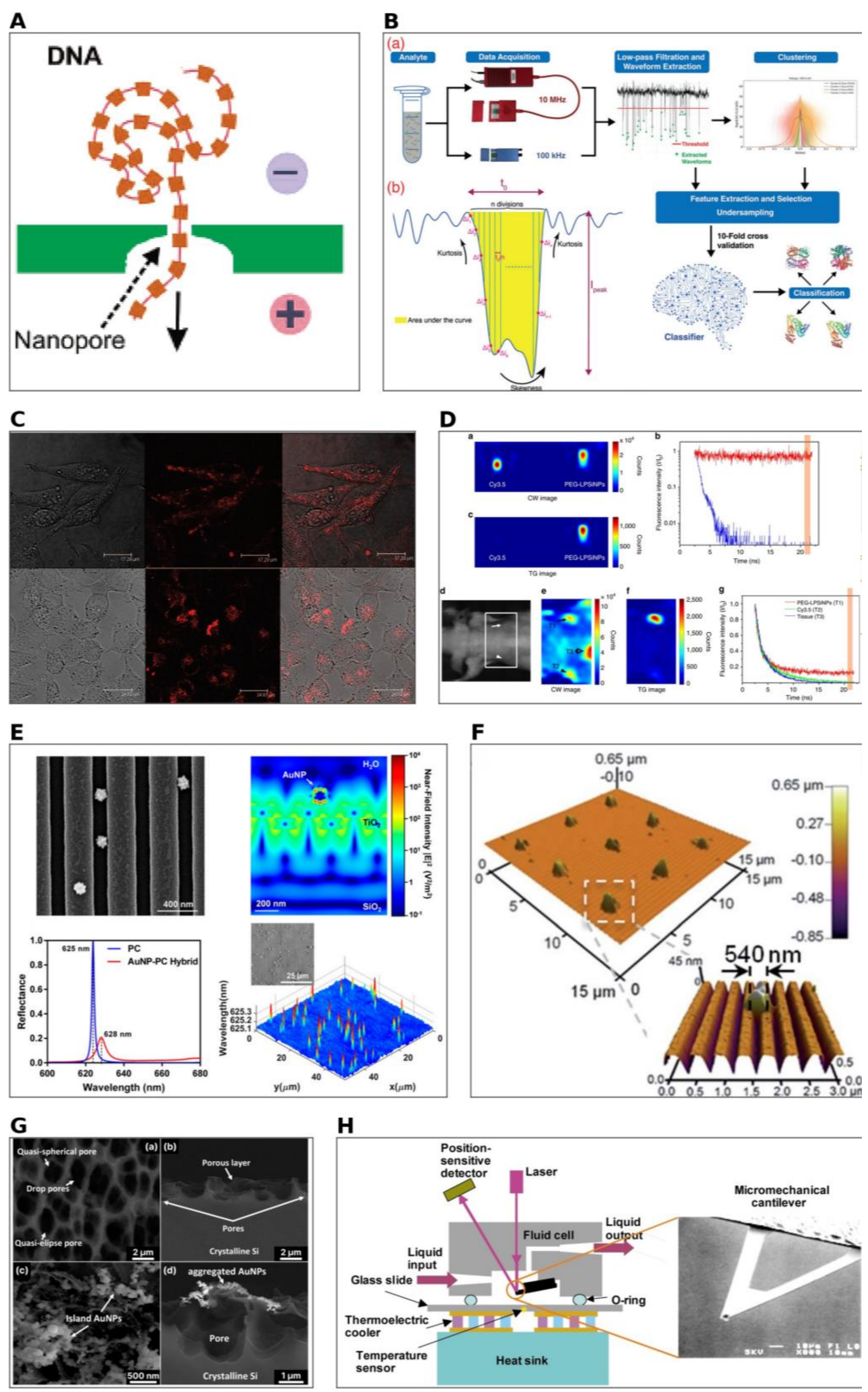

**Figure 8.** Emerging silicon biosensing platforms. (A) Single-stranded DNA translocation events recorded through a solid-state silicon-nitride nanopore; adapted from Fologea et al. [59] with permission from the American Chemical

Society, copyright 2005. (B) Machine-learning-assisted protein identification from solid-state nanopore signals; reproduced from Dutt et al. [66] under Creative Commons Attribution License (Wiley-VCH, 2023). (C) Biocompatible luminescent silicon quantum dots imaging cancer cells; adapted from Erogbogbo et al. [54] with permission from the American Chemical Society, copyright 2008. (D) In vivo time-gated fluorescence imaging using biodegradable luminescent porous silicon nanoparticles; reproduced from Gu et al. [56] under Creative Commons Attribution License (Springer Nature, 2013). (E) Photonic resonator absorption microscopy for digital, single-base-selective microRNA detection on a silicon photonic-crystal surface; adapted from Canady et al. [97] with permission from PNAS, copyright 2019. (F) Single-nanoparticle detection by photonic crystal enhanced microscopy; adapted from Zhuo et al. [96] with permission from the Royal Society of Chemistry, copyright 2014. (G) Morphology and field-enhancement of black silicon SERS substrates; reproduced from Padrez and Golubewa [98] under Creative Commons Attribution License (MDPI, 2024). (H) Microcantilever array used for label-free PSA bioassay; adapted from Wu et al. [86] with permission from Springer Nature, copyright 2001.

Silicon quantum dots and luminescent silicon nanoparticles are at an even earlier stage of translation, but they are conceptually rich platforms. A useful entry point into this field is the 2023 review by Zhang et al. [52]. To illustrate the current direction of SiQD development, Sebastian et al.'s water-soluble SiQD fluorescence immunoassay probe for C-reactive protein [53] provides an immediately relevant example. As illustrated in Figure 8(C), the use of biocompatible luminescent silicon quantum dots for cancer-cell imaging was established in earlier work by Erogbogbo and colleagues [54]. Park et al.'s biodegradable luminescent porous-silicon nanoparticles for in vivo applications [55] remain one of the clearest demonstrations that silicon-based emitters can survive in biological tissue and undergo productive biodegradation. As illustrated in Figure 8(D), Gu and colleagues extended this material platform to in vivo time-gated fluorescence imaging [56], providing one of the first compelling whole-animal readouts in the field.

Photonic-crystal-enhanced biosensing on silicon is closely related to silicon photonics, but it operates within a distinct device geometry. The BIND label-free assay system developed by Cunningham et al. [92], quantum-dot fluorescence enhancement on photonic-crystal surfaces demonstrated by Ganesh et al. [93], a general photonic-crystal biosensor method for discovering inhibitors of protein–DNA interactions reported by Chan et al. [94], label-free smartphone-based biodetection reported by Gallegos et al. [95], photonic-crystal-enhanced microscopy for single-nanoparticle detection by Zhuo et al. [96], as illustrated in Figure 8(F), and photonic-resonator absorption microscopy for digital-resolution microRNA detection with single-base selectivity by Canady et al. [97], as illustrated in Figure 8(E), collectively trace an arc from large-format screening to single-molecule digital readout.

Black silicon SERS represents one of the newest directions discussed in this review and is among the youngest silicon nanostructure platforms for biosensing. As illustrated in Figure 8(G), the field has recently been consolidated by the 2024 review of black-silicon SERS biosensors by Padrez and Golubewa [98]. Earlier work on surface-plasmon-enhanced broadband spectrophotometry on black silver by Xu et al. [99], together with the chemically non-perturbing SERS detection of catalytic reactions on black silicon by Mitsai et al. [100], provides the broader plasmonic context and demonstrates that the black-silicon substrate can play an active functional role rather than serving merely as a passive support.

The emerging platforms discussed above share a common structural challenge: each is promising for a relatively narrow set of biosensing applications, yet none has fully justified the manufacturing investment required for an industry-scale translational push within its own commercial niche. Among these platforms, solid-state nanopores appear closest to commercialization, because they align with an already active DNA and RNA sequencing market. By contrast, SiQDs and black-silicon SERS remain further away from translation, partly because their value propositions substantially overlap with established fluorescence and Raman ecosystems. We therefore anticipate that, over the next five years, the central question in each of these subfields will shift from “Can this platform sense X?” to “Can this platform solve a clinically important problem in a way that no competing platform can, and can it do so reproducibly at industrial scale?”

### 3.6. Cross-platform comparisons

Cross-platform comparison continues to be treated as a research problem in its own right within parts of the silicon biosensor literature. Two main lines of argument stand out. The first concerns when CMOS compatibility actually matters. For SiNW-FETs, CMOS compatibility is primarily an argument about manufacturability and scale-up; the underlying device physics is essentially the same, as discussed in Tran's review [16]. For silicon photonic resonators, the argument remains broadly valid as well, because telecom-grade transceivers already rely on the same deep-UV lithography processes. By contrast, CMOS compatibility is more contested for porous silicon for two reasons: first, anodization is not a standard CMOS fabrication step; and second, unless the porous region is otherwise isolated, the resulting morphology can be damaging to neighboring CMOS structures.

The second line of argument concerns whether these platforms should be combined or treated as competing alternatives. Several recent studies argue for hybrid devices that integrate different readout modalities, for example by pairing charge-based readout with optical or single-molecule readout on the same chip. Plasmonic-augmented silicon photonics, combined SiNW-FET-microring designs, and photonic-crystal-enhanced absorption microscopy all point in this direction [97]. The key trade-offs that we consider most relevant for working research groups choosing among these platforms are summarized in Table 1.

**Table 1.** Critical comparison of mature and emerging silicon biosensor platforms across transduction mechanism, demonstrated sensitivity range, strongest validated application, most acute current limitation, and translation readiness. The table is meant to be read alongside the running text rather than as a stand-alone ranking; for any given analyte and matrix the optimal platform is likely to be different from any other.

| Platform | Transduction | Best demonstrated LoD range | Strongest validated application | Most acute current limitation | Translation readiness | Key refs |
|---|---|---|---|---|---|---|
| SiNW / nanoribbon BioFETs | Surface-potential modulation of subthreshold current | fM nucleic acids in buffer; pg/mL proteins in spiked serum | Multiplexed protein and pathogen detection in low-salt buffer | Debye screening; sensor-to-sensor variability; drift in undiluted plasma | Medium; one commercial CMOS-FET sequencer (Ion Torrent) | [17–20, 23, 28, 29, 33] |
| Porous silicon (optical) | Effective optical thickness shift (Fabry–Pérot, RIFTS, photonic crystal) | nM–pM proteins; single-bacterium for some target classes | Real-time bacterial detection in food and water matrices | Surface stability in physiological buffers; pore accessibility tradeoffs | Medium; food-safety and POC demonstrations | [10, 36–46] |
| Porous silicon (electrochemical) | Faradaic current / impedance change | Sub-fM bacterial RNA in stabilized formats | Bacterial RNA and small-molecule electrochemical detection | Long-term electrode stability and drift | Early translational | [47–51] |
| Silicon photonic resonators | Resonance wavelength shift with refractive index | ng/mL proteins in serum; multiplex viral antigens | Multiplexed clinical biomarker quantitation | Thermal drift; packaging; off-chip spectrometer cost | High at chip level; mid at system level | [68–84] |
| Silicon / SiN nanopores | Ionic current blockade during translocation | Single-molecule DNA; protein identification with ML | Single-molecule DNA / protein sensing | Pore stability; fabrication yield; throughput | Low to medium; tied to sequencing market | [57–67] |
| Si quantum dots / luminescent Si NPs | Photoluminescence, FRET, lifetime | Sub-ng/mL biomarkers in immunoassay | In vivo imaging; fluorescence immunoassay | Surface chemistry; quantum-yield stability; biocompatibility data | Low; few clinical demonstrations | [52–56] |
| Photonic crystal / black Si SERS | Enhanced fluorescence / Raman or resonator absorption | aM–fM microRNA at single-base selectivity | Digital-resolution miRNA; label-free smartphone readout | Substrate uniformity; signal calibration | Medium for photonic crystal; low for SERS | [92–100] |
| MEMS cantilevers | Surface-stress bending or resonant-frequency shift | ng/mL proteins | Surface-stress immunosensing in air | Damping in liquid; integration with CMOS readout | Low; niche only | [85–91] |

For readers, three questions are worth asking before accepting record-low LoD values reported in any new paper. What was the sample matrix? What calibration scheme was used? How many sensors were measured, and what measurement statistics were reported? Only when all three questions are answered can we assess whether the device's performance is genuinely promising and whether it fairly surpasses existing alternatives.

It may be misleading, and in some cases harmful to the field, to declare a single "winning" platform. Different platforms are best suited to different combinations of analytes, matrices, deployment settings, and price points. The field would benefit more from head-to-head comparisons of platforms with specific applications than from isolated platform demonstrations alone. The most useful comparisons among silicon biosensor platforms should therefore not focus solely on the highest sensitivity or lowest LoD values, but rather on which platforms can define the most accurate, reliable, reproducible, and broadly executable bench protocols for each application-specific use case.

# 4. Challenges, Conclusions, and Outlook

## 4.1. Challenges and limitations

Several challenges have emerged in the development of silicon biosensors over the past three decades. The first is Debye screening in BioFETs, which remains the most widely discussed – and perhaps the least well understood – bottleneck in the field. For conventional operation in physiological salt concentrations, the classic limit described by Knopfmacher and colleagues [22] still applies. The two most credible workarounds are Elnathan's biorecognition-layer engineering strategy [24] and Kulkarni and Zhong's high-frequency operation approach [25]. Extending the high-frequency approach, radio-frequency modulation of silicon nanowire FETs has been shown to enhance detection sensitivity through flexoelectric resonance while perturbing the electrical double layer to reduce Debye screening in high-ionic-strength media, enabling direct biomarker detection without sample dilution [101, 102]. However, very few studies have progressed from spiked buffer experiments to demonstrations in undiluted patient blood while retaining both sensitivity and selectivity. The absence of such validation should be regarded by reviewers as an important warning signal.

The second challenge is sensor-to-sensor variability. Tintelott's analysis of process variability in top-down silicon nanowire fabrication [15] and Zafar's minimal-variation devices [28] make clear that, without strict process discipline, wafer-scale BioFET sensitivity cannot yet be guaranteed. Chen et al.'s calibration framework [29] further reframes this issue as a measurement problem rather than a fabrication-only problem. Across platforms, the most credible recent demonstrations do not merely report one-off limits of detection; they also provide device-to-device statistics. Porous-silicon stability and surface fouling follow closely behind. The field has developed much of the necessary chemistry, as suggested by Mariani and colleagues' layer-by-layer approach [44] and the carbon-stabilized electrochemical platforms reported by Guo and colleagues [49, 50]. However, it still lacks a standardized fouling protocol against which new chemistries can be benchmarked and on which regulators can rely.

The third challenge is integrated readout. A wafer-scale sensor may be compact, but if the spectrometer or associated instrumentation occupies a benchtop, the overall system is not portable, and the cost per test remains dominated by the external hardware. Several strategies are being pursued to reduce instrument cost and footprint, including the integration of filter banks and spectrometers onto the chip itself [79, 81], the deployment of LED-pumped active resonators [84], and the emergence of smartphone-based optical biodetection on silicon [95]. Optical silicon biosensors are likely to remain bench-top instruments until these approaches mature. Table 2 summarizes how each study addresses these three challenges and highlights the most visible remaining gaps.

**Table 2.** Recurring bottlenecks in silicon biosensor translation and where the literature is meaningfully addressing them. The third column captures the most credible technical responses we have found; the fourth captures the gaps that still keep these platforms out of routine clinical use.

| Bottleneck | Why it matters | Current strategies in the literature | Where the gap remains | Refs |
|---|---|---|---|---|
| Debye screening | Limits BioFET sensitivity in physiological salts | Biorecognition-layer engineering; high-frequency operation; fragment-Fab linkers | Few translation-ready chemistries; sparse data in undiluted blood | [22, 24, 25, 27] |
| Sensor-to-sensor variability | Confounds quantitative thresholds and external calibration | Calibration frameworks; tight CMOS process control | Lack of cross-vendor calibrators and reference materials | [15, 16, 28, 29] |
| Pore wetting and accessibility | Limits mass transport into the high surface area of porous silicon | Microfluidic integration; flow-through architectures | No standardized hydrodynamic benchmark across labs | [40, 43, 45] |
| Surface fouling | Reduces effective LoD in plasma and serum | Layer-by-layer chemistries; zwitterionic and antifouling polymers | No widely used reference fouling protocol | [44, 80] |
| Spectral readout and packaging | Sensor chip is cheap; spectrometer / thermal control is not | On-chip filter banks; integrated NIR spectrometers; LED-pumped resonators | Cost-per-test still spectrometer-limited | [79, 81, 84] |
| Real-matrix validation | Most LoDs are in buffer, not in patient samples | Recent serum / plasma serial demonstrations; multiplex viral assays | Few six-month, multi-operator clinical comparison studies | [29, 30, 31, 78, 82] |
| Manufacturing reproducibility for emerging platforms | Limits commercial scale-up of nanopores, SiQDs, SERS | Femtosecond-laser nanopore fab; standardized SiQD synthesis | Most demonstrations remain lab-scale | [52, 67] |

If a single most important challenge must be selected from Table 2, it is real-matrix validation. Sensitivity in the buffer can be improved through better chemistry, but clinical credibility cannot be established without testing in clinically relevant samples. The field would benefit far more from one rigorously executed, multi-operator, multi-month, blinded study of a single silicon biosensor in real patient samples than from another hundred femtomolar-limit-of-detection demonstrations in PBS. Funding agencies and journals should help accelerate this transition by prioritizing real-matrix studies and by requiring authors to report appropriate matrix controls, sample statistics, and device-to-device variability. No new physics is required; what is needed are new norms.

### 4.2. Conclusions

Considered category by category, silicon nanostructures are not necessarily the highest-Q resonators, the brightest emitters, the thinnest membranes, or the most sensitive electrodes currently available. What silicon nanostructures offer instead is a universal, CMOS-compatible platform on which diverse sensing elements can be built together on the same wafer, with coordinated design, shared processing, compatible chemistries, and high levels of integration. This distinctive combination of attributes makes silicon biosensors highly competitive with other transduction mechanisms and sensing modalities and positions them as a promising platform for the next decade of biosensor signal translation.

The highest-impact direction identified in this review is interface engineering. Silicon BioFETs are not primarily limited by intrinsic transduction, but rather, they are limited by the processes occurring within the first 1–5 nm above the gate oxide, where water, ions, biomolecules, analytes, and Debye screening collectively determine the electrical signal ultimately measured between the source–drain electrode pair. Three studies – the biorecognition-layer engineering reported by Elnathan et al. [24], the high-frequency operation proposed by Kulkarni and Zhong [25] and extended by Liu et al. [101], and the calibration framework developed by Chen et al. [29] – all point to the same conclusion: the signal generated by silicon nanostructures is real, but its interpretation is determined by the interface. A similar conclusion applies to porous silicon, where the field has shifted from pore design toward pore conditioning, as reflected in the layer-by-layer chemistry of Mariani et al. [44], the carbon-stabilized electrochemical platforms of Guo and colleagues [49, 50], and the antifouling, microfluidically integrated devices reported by Awawdeh [43]. In silicon photonics, the biorecognition layer has likewise become a rate-limiting factor, as shown by Puumala et al. [80], who emphasize its central role in determining multiplexed assay performance. This cross-domain importance of the interface was anticipated in the 2002 review of biologically sensitive FETs by Schöning and Poghossian et al. [3], and the broader silicon-biosensor literature from 1970 [1] to 2026 [68] continues to confirm it.

Two overlooked truths sit underneath this conclusion. First, the major bottleneck does not lie in the silicon platform itself, but in the interface and the readout architecture. Second, the most consequential work in the field over the past five years has not been the demonstration of increasingly novel transduction mechanisms, sensing modalities, or ultralow limits of detection in conventional PBS buffers; rather, it has been the rigorous validation of integrated platforms in real patient matrices. Funding agencies and journals should therefore place greater emphasis on these two directions, prioritizing interface and readout engineering as well as statistically robust validation in clinically relevant samples.

### 4.3. Outlook

To look forward to the future of the field, we envision four major directions for the coming decade. The first is readout integration: moving filter banks, spectrometers, signal-processing electronics, and related components onto the chip, or at least co-packaging them with the sensor in a compact module. Several efforts have already advanced this goal, including the integration of spectrometers directly onto silicon photonic platforms [79, 81], the development of LED-pumped active resonators [84], and the emergence of smartphone-based optical biodetection using silicon-based devices [95]. We anticipate that, within the next decade, integrated readout will become one of the most critical determinants of whether silicon photonic biosensors can succeed in clinical settings – potentially even more decisive than incremental improvements in sensor-chip sensitivity.

The second future direction is to treat variability and calibration as first-class challenges rather than afterthoughts. This priority has been articulated from the fabrication perspective through Tintelott's analysis of process variability [15] and Zafar's minimal-variation devices [28], and from the measurement perspective through the calibration framework developed by Chen et al. [29]. We expect that the most influential silicon biosensor studies in the next five years will be those that report device-to-device variability, long-term drift, and cross-laboratory reproducibility as routinely as they currently report improved sensitivity.

The third future direction is the maturation of emerging silicon-based sensing platforms. For solid-state nanopore biosensing, the field is moving beyond simple physics demonstrations toward machine-learning-assisted protein identification, in which amplification-based assays are complemented by single-molecule resolution and probabilistic readout [66, 67]. Based on the in-vivo studies by Park [55] and Gu [56], silicon quantum dots are likely to become increasingly useful as in-situ optical labels and tracers in tissues and cells, particularly as biomarker-oriented applications begin to consolidate, as exemplified by Sebastian's CRP probe [53]. Rather than replacing silicon photonics, newer photonic-crystal-enhanced and black-silicon SERS approaches will likely develop alongside it, especially in applications where spectroscopic identification, rather than simple concentration measurement, is required [98-100].

The fourth future direction is the expanded use of silicon biosensors in traditional application domains beyond the clinical laboratory. Two domains appear particularly well matched to silicon-based sensing. The first is environmental and food-safety monitoring, where silicon-photonic and porous-silicon platforms have shown promising validation for the detection of bacteria, viruses, and toxins [41, 42, 82]. The second is wearable and point-of-care diagnostics, where MEMS cantilevers [91] and miniaturized Si Quantum Dots (SiQD)-based assays [52, 53] may gain traction before achieving broader adoption in centralized clinical laboratories.

For researchers entering this field today, our recommendation is not to focus primarily on inventing yet another silicon nanostructure. Instead, the more impactful path is to take a mature silicon platform, deploy it in a real sample matrix, rigorously characterize and validate its performance, co-design the device architecture and interface chemistry, report not only improved sensitivity but also calibrated variability and statistical reproducibility, and integrate the sensing and readout components into a complete system capable of replacing large benchtop instrumentation. This, we argue, is the future direction through which silicon biosensors are most likely to succeed in clinically relevant applications.

We also note several developments that are unlikely to occur in the near term. It is unlikely that a novel silicon nanostructure will emerge and replace all of the major platforms discussed above. It is also unlikely that silicon biosensors will replace mature clinical immunoassays at scale within the next five years. Within a decade, a silicon biosensor may plausibly become a consumer wearable device comparable in impact to the glucose monitor, but only if interface stability and long-term operational robustness improve substantially. This outlook is not intended to be pessimistic; rather, it reflects the view that the field should prioritize problems whose solutions are both consequential and realistically within reach.

## Author Contributions

A.L. conceived the scope of the review, conducted the literature search, drafted the manuscript, designed and prepared all figures and tables, derived the governing equations, and serves as the corresponding author. J.C. contributed to the bioconjugation chemistry and molecular-recognition sections, assisted with the preparation and critical review of figures and tables, and edited the manuscript for clarity and technical accuracy. Z.S. contributed to the fabrication and surface-functionalization sections, assisted with reference curation for the porous-silicon and silicon-photonics subsections, and provided critical feedback on the figures and tables. J.S. supervised the project, shaped the framing of the challenges, conclusions, and outlook sections, provided expertise on the photonic and optical-sensing content, edited the manuscript for technical accuracy and clarity. During preparation, the authors used AI-based tools solely to assist with proofreading, correcting typographical errors, and polishing language. All authors discussed the scientific content and analysis, reviewed the manuscript, and approved the final version for submission.

## Conflicts of Interest

The authors declare no competing financial interests. No author has a consulting, advisory, employment, or equity relationship with any of the device manufacturers, instrument vendors, or commercial assay providers mentioned in this review.

## Funding Statement

This review received no external funding. No grant agency, company, or institution provided dedicated funding or support for the preparation of this article.